# Evolution of Microstructure and Relaxor Ferroelectric Properties in $(La_zBa_{1-z})(Ti_{0.80}Sn_{0.20})O_3$

R. Kumar[1,2], K. Asokan[3], S. Patnaik[2] and B. Birajdar[1]

[1]*Special Centre for Nano Sciences, Jawaharlal Nehru University, New Delhi, India*
[2]*School of Physical Sciences, Jawaharlal Nehru University, New Delhi, India*
[3]*Mateials Science Division, Inter University Accelerator Centre, New Delhi, India.*

## Abstract

We report a study on a series of lanthanum doped barium stannate-titanate (LBTS) ceramics towards correlating their microstructure with ferroelectric properties. The samples were prepared by solid state reaction technique and the XRD analysis indicated predominantly single phase microstructure. Scanning Electron Microscopy revealed that while a low La doping of 1 atomic % yielded inhibited grain growth in 1LBTS, higher amounts of La doping yield larger leafy grains. This peculiar grain growth and microstructure evolution could be attributed to the combination of the following two competing factors: (i) suppression of O vacancies due to La doping leading to grain growth inhibition and (ii) enhancement of mechanism leading to nucleation and growth of 2-D (leafy) grains for La doping higher than 2 atomic % due to eutectic phenomena. Dielectric permittivity measurements indicate that addition of La enhances diffused phase transition in BTS, but its frequency dependence is observed only at 2 atomic % La doping. This signature relaxor behaviour and the increase in the intensity of a particular Raman peak with La doping, as well as non-saturated PE hysteresis loops confirm the presence of polar nano-regions for 2 and 3 atomic % La doped BTS. However significant enhancement in dielectric permittivity could not be achieved due to reduced grain size.

1. Introduction

Encompassing several potential applications, the barium titanate ($BaTiO_3$) based relaxor ferroelectric materials have attracted considerable attention in the recent past [1]. In particular, relaxors are potential materials for capacitors that generally require high and frequency independent dielectric constant over a broad range of temperature [2]. Apart from being environment-friendly, the properties of $BaTiO_3$ (BTO) ceramics are tunable to such requirements by suitable substitutions at Ba and Ti sites. By definition, relaxor ferroelectrics are characterized by broad maxima in the temperature dependent dielectric permittivity that shifts to higher temperature with increasing frequency [3-4]. This phenomenon was first evidenced in $Ba(Ti_{1-x}Sn_x)O_3$ by Smoleskii et al. (in 1954) and is currently ascribed to the development of polar nano regions (PNRs) [5-6]. This is in contrast to macroscopic polarized domains in pure $BaTiO_3$ that exhibit sharp dielctric peak at paraelectric (cubic) to ferroelectric (tetragonal) phase transition ($T_c$~403 K) [7]. Several reports substantiate a diffused phase transition in $Ba(Sn_xTi_{1-x})O_3$ that achieves a complete relaxor characteristics for x ~ 0.30 [8]. On the other hand, A site substituted $La_zBa_{1-z}Ti_{1-z/4}O_3$ exhibits record high dielectric constant (~ 25,000) alongwith exceptional electromechanical properties [9-10]. This substitution of $La^{3+}$ in place of $Ba^{2+}$ is linked to the creation of B-site vacancies for charge compensation. Thus, heterovalent as well as homovalent substitutions at A site (La) and B site (Sn) can be used towards tuning relaxor properties by controlling microstrucutre and phase segregation [9]. In this communication, we report the microstructure formation in $(Ba_{1-z}La_z)(Ti_{0.80}Sn_{0.20})O_3$ using scanning electron mciroscopy (SEM) and have tried to correlate it with relaxor ferroelectric properties. Furthermore, co-doping of Sn and La in Barium Titanate has been investigated using Raman Spectroscopy to study the impact of La addition on relaxor behavior.

2. **Experimental Technique**

The ceramic samples were prepared by conventional solid state reaction technique with starting materials $BaCO_3$, $La_2O_3$, $TiO_2$ and $SnO_2$ with purity $\geq$ 99% were purchased from Sigma Aldrich. They were mixed according to the stoichiometric ratio to obtain nominal compositions of $(Ba_{1-z}La_z)(Ti_{0.80}Sn_{0.20})O_3$ (z = 0.0, 0.01, 0.02 and 0.03). In this manuscript these compositions are denoted as BTS, 1LBTS, 2LBTS, 3LBTS respectively. After 35 hrs electric grinding, the mixed powder was pressed into pellets of diameter 10 mm and thickness about 1 to 1.5 mm. Optimum calcination conditions were found to be 4 hrs at 1473 K for BTS and 8 hours at 1523 K for all LBTS samples. The calcined pellets were then reground and pressed into pellets of 10 mm diameter. Optimum sintering conditions were found to be 3 hrs at 1653 K for BTS and 4 hrs at 1673 K for LBTS samples except for 3LBTS which was sintered for 9 hrs. All sintering was done in presence of air in alumina crucible that resulted in highly densified ceramics. The XRD patterns of the samples were obtained with *Rigaku* diffractometer using *Cu-$K_\alpha$* radiation. The diffractometer was operated at voltage 30 kV and current 20 mA with the scanning speed of 4°/min for the diffraction angle 2θ in the range of 20-80°. Secondary electron images of the gold coated pellets were acquired using ZEISS EV40 scanning electron microscope (SEM). Raman spectra were acquired at room temperature over a wavenumber range of 50 – 1200 cm$^{-1}$ using Raman Spectrometer (Enspectr Enhanced spectroscopy, USA) with a green excitation light source of 532 nm line. The dielectric measurements were done using Agilent E4980A Precision LCR meter. It was carried out as a function of frequency of the applied ac field (range 25Hz–1MHz) under zero dc bias over a temperature range from 80 K to 350 K. The temperature was controlled with accuracy 1 K by using a programmable temperature controller (Lake Shore 325). To prevent electrical breakdown along the edge of the pellets and also to avoid fringing effects, only one side of the pellets was fully coated with silver paste. The other side of the pellets was partially (70-80% diameter) coated. The real part of dielectric constant ($\epsilon$) was calculated from the relation $\epsilon = Ct/\epsilon_0 A$ where C, $\epsilon_0$, t and A are the capacitance, permittivity of free space, thickness and coated area of the sample respectively. The Polarization versus Electric field (PE) loop for all the compositions is recorded at a frequency of 50 Hz by applying a maximum electric field of 30 kV/cm at room temperature.

## 3. Results and Discussion

Room temperature X-ray powder diffraction pattern of $(Ba_{1-z}La_z)(Ti_{0.80}Sn_{0.20})O_3$ with (z=0.00,0.01,0.02 and 0.03) sintered ceramics are shown in Fig.1. XRD patterns of BTS and LBTS samples are in agreement with JCPDS Card No. 62−0356. Further, in agreement with existing reports, BTS and LBTS samples show cubic structure with space group Pm3m [11-12]. Extra peaks due to impurities were not found in the XRD patterns, indicating predominant single phase formation for all the prepared samples. Inset of Fig.1 shows the enlarged view of XRD pattern in the range (2θ = 37—47°) to depict the right shifting of peaks with La addition in BTS. Evidently, lattice parameter and unit cell volume decreased with increase of lanthanum doping in BTS. This is in accordance with smaller ionic radius of $La^{3+}$ (1.36 Å) compared to $Ba^{2+}$ (1.61 Å) which are reflective of successful substitution of La at the Ba site. The XRD pattern also indicates that the FWHM of the most intense diffraction peak (2θ = 31.5°) is getting broader with increase in La doping. This is indicative of a decrease in crystalline size with increased La addition.

Secondary electron images of $(Ba_{1-z}La_z)(Ti_{0.80}Sn_{0.20})O_3$ with (z=0.00,0.01,0.02 and 0.03) ceramic samples are shown in Fig 2 (a-d). Pure BTS sample (Fig. 2 (a)) reveals 3-dimensional

(3-D) grains with a size of 5-30 micro-meters. Many of the larger grains show terraced surface. Addition of La in BTS is found to inhibit grain growth. 1LBTS sample (Fig. 2 (b)) reveals grains of 0.5 to 2-micro-meters. 2LBTS sample (Fig. 2 (c)) clearly reveals a bimodal microstructure consisting 3-D submicron sized grains together with leafy (2-D) grains of the size of about 2-micrometers. Thus for 2 atomic % La doping, the growth of 3-D grains continues to be inhibited but nucleation and anomalous growth of 2-D (leafy) grains is observed. Thus 2 atomic % appears to be critical La doping level for the nucleation and growth of 2-D leafy grains. 3LBTS sample (Fig. 2 (d)) seems to be dominated by an extremely compact arrangement of leafy grains with a size of 2-5 micro-meters, which might be attributed to the higher La doping level as well as to the longer sintering time used for this sample. Thus, while a low La doping of 1 atomic % yielded inhibited grain growth in 1LBTS, higher amounts of La doping yield larger leafy grains in 2LBTS and 3LBTS. Similar results- grain growth inhibition at low La doping and enhanced grain growth at high La doping has been observed in La doped $Ba(Zr,Ti)O_3$ samples [13]. A systematic observation of microstructure of La doped BTS samples and mechanism of its formation has not been reported in the literature. Depending upon synthesis conditions, different competing doping mechanisms have been proposed. Formation of oxygen vacancies due to the addition of $SnO_2$ in $BaTiO_3$ has been reported [14]. Similarly, presence of unavoidable acceptor impurities has also been reported to yield O vacancies [15]. Oxygen vacancies are also known to result in samples annealed at high temperatures (> 1573 K) due to the evaporation of O [9]. The formation of large grains with terraced surfaces in BTS sample (Fig. 2 (a)) could be attributed to the enhanced mass transport via grain boundaries due to these O vacancies. With doping of La, the formation of O vacancies is suppressed [12] which might explain the inhibited grain growth at low La doping. For high La doping, addition of significant amount of $La_2O_3$ in the precursor powder is likely to result in the lowering of melting point due to eutectic phenomenon [16]. The associated enhancement in the diffusion and mass transport explains the enhanced grain growth at high La doping. It has been reported that if the sintering temperature is not high enough, the grain surfaces may be approximated as atomically smooth structures and grain growth occurs by 2-D nucleation and lateral growth mechanism [16]. We propose that the mole fractions and sintering temperature of 1673 K for 2LBTS and 3LBTS samples are suitable for 2-D nucleation and lateral growth leading to formation and growth of leafy grains. In this way the peculiar grain growth pattern - inhibited grain growth at low La doping but enhanced and leafy grain growth for high La doping can be attributed to the combination of the following two competing factors: (i) suppression of O vacancies due to La doping leading to grain growth inhibition and (ii) enhancement of mechanism leading to nucleation and growth of 2-D (leafy) grains for La doping higher than 2 atomic % due to eutectic phenomena.

Room temperature Raman spectra of $(Ba_{1-z}La_z)(Ti_{0.80}Sn_{0.20})O_3$ with (z=0.00,0.01,0.02 and 0.03) are shown in Fig. 3. In a perfect cubic structure Raman modes are forbidden but broad Raman peaks are none the less obtained which is attributed to the disorder in the A or B sub-lattice [17] and to distorted oxygen octahedra [18-19]. The Raman spectra show six prominent modes and one anti resonance dip at 117, 192, 281, 506, 735, 836 and 141cm$^{-1}$ respectively. These peaks, barring the one at 836 cm$^{-1}$, match well with the reported data on Sn doped BaTiO$_3$ ceramics [20]. In particular, the intensity under the $A_{1g}$ Transverse Optical mode at 506 cm$^{-1}$ was correlated with the formation and dynamics of polar nano-regions [20]. With increase in La doping the intensity under this peak increases indicating enhanced formation of polar nano regions. In the literature the peak close to 735 cm$^{-1}$ in Sn doped BTO samples have been interpreted differently by various authors. While Baskaran et al. [21] and Pokorny et al. [19]

have associated this peak to the ferroelectric breathing of $BO_6$ octahedra (tetragonal distortion) in the Sn doped BTO and pure BTO respectively, Kumar et al. [20] observed that it resembles the $A_{1g}$ mode of classic relaxors associated with the breathing of $BO_6$ octahedra. It is worthwhile to compare it with the case of BZT, in which the peak at ~715 cm$^{-1}$ is associated with ferroelectric breathing of $BO_6$ octahedra while the peak at ~780 cm$^{-1}$ is associated with a $A_{1g}$ stretching mode of $BO_6$ octahedra which becomes Raman active because of the presence of dissimilar B site species at the centre of octahedra [19, 21-22]. On increasing La doping in BZT, the peak intensity under $A_{1g}$ mode increases [18]. In the present work however, the most noticeable change due to La doping in BTS is the appearance of peak at 839 cm$^{-1}$. It has been reported that La doping in pure BTO also results in the appearance of the $A_{1g}$ relaxor phase peak at 840 cm$^{-1}$ which was attributed to the Ti vacancies created at B sites due to La doping at A sites [21]. In view of the preceding discussion it is proposed that the peak at 735 cm$^{-1}$ in BTS and LBTS is associated with the ferroelectric breathing of tetragonal $BO_6$ octahedra and the peak at 839 cm$^{-1}$ is associated with the $A_{1g}$ stretching mode of $BO_6$ octahedra which is characteristic of relaxor phase. This raises the question as to why the $A_{1g}$ mode is absent in BTS but present in BZT. The explanation for this could be the similar ionic radii of $Sn^{4+}$ (0.71 Å) and $Ti^{4+}$ (0.68 Å); the ionic radius of $Zr^{4+}$ on the other hand is much larger (0.79 Å).

Temperature dependent dielectric permittivity ($\epsilon$) and loss factor (tan δ) at different frequencies of sintered $(Ba_{1-z}La_z)(Sn_{0.20}Ti_{0.80})O_3$ with (z= 0.00, 0.01, 0.02 and 0.03) ceramics are shown in Fig. 4. A significant increase in full width at half maxima (FWHM) along with decrease in maximum dielectric permittivity ($\epsilon_m$) and Curie temperature ($T_c$) with increase of La doping has been reported [11]. The extent of relaxor ferroelectricity characteristics is parameterized by the *exponent γ* in the modified Curie-Weiss law; [24-25]

$$\frac{1}{\epsilon} - \frac{1}{\epsilon_m} = \frac{(T-T_m)^\gamma}{C}, (T > T_m) \quad (1)$$

Where $T_m$ is the temperature corresponding to the dielectric permittivity maximum, C is the Curie constant, $\epsilon_m$ is the peak dielectric permittivity at temperature $T_m$ and exponent γ reflects the nature of relaxor phase transition. Inverse dielectric constant plots as function of temperature at 100 kHz are shown in Fig. 5. The inset in each panel shows plots of $\ln(1/\epsilon - 1/\epsilon_m)$ verses $\ln(T-T_m)$ for all compositions, the slopes of which determine the exponent γ. The experimental data were fitted to equation (1) and the best-fit values for $T_m$, $\epsilon_m$, and γ were obtained at frequency 100 kHz. Further, $\delta T_{relax} = T_m$ (1 MHz) — $T_m$ (1 kHz) and Full width at half maxima (FWHM) were also determined. All these values are summarized in Table 1 that is discussed below.

As seen in Fig. 4 (a and b), BTS and 1LBTS samples do not show noticeable shift in dielectric peak transition temperature ($T_m$) with increasing frequency as $\delta T_{relax}$ are found to be 0 K and 4 K respectively. It is however worth noting that even 1 atomic % of La doping in BTS has resulted in marked increase in the FWHM of the phase transition (Table 1) and the value of maximum dielectric constant is less sensitive to the applied frequency (Fig. 4 (b)). This negligible frequency dispersion and broad peak is signature of ferroelectricity with *diffused* phase transition. On the other hand, significantly higher dielectric dispersion, $\delta T_{relax}$= 10 K and 16 K, is observed for La doping (z= 0.02 and 0.03 respectively) in Fig. 4 (c, d). Thus we conclude that 2 atomic % of La doping is able to induce relaxor behaviour in BTS20 sample. In comparison $\delta T_{relax}$ of 10 K was obtained in BTS15 sample for a La doping of 3 atomic % [11]. In contrast, it is reported that in the absence of La doping, the Sn doping would have to be increased to 30 atomic % to induce relaxor behaviour in pure Barium Titanate [26]. As has been well documented in the literature [12-13] substitution of $La^{3+}$ (ionic radius 1.32 Å) in place of $Ba^{2+}$

(ionic radius 1.61 Å) introduces Ti vacancies or $Ti^{3+}$ in the lattice. This disorder introduces random electric fields and strain because of which polar nano regions are induced. The flipping or the breathing of polar nano-regions is considered to be responsible for the relaxor properties [27].

From equation 1, it is clear that in normal ferroelectric transition, γ=1 while for complete relaxor behaviour ~ 2 and an incomplete-diffused phase or relaxor phase transition, $1 \leq \gamma \leq 2$ [28]. It is observed that as La addition increases from z=0.00. 0.01, 0.02 and 0.03, the γ increases from 1.42, 1.88, 2.017 and 2.018 respectively. This conclusively proves that La addition onto BTS drives the system into a robust relaxor phase. Although the broadening (with respect to temperature) of the phase transition is achieved by La doping, the maximum dielectric constant has actually decreased. This decrease could be attributed to the increased volume fraction of grain boundaries, due to decrease in grain size (Fig. 2). Similar decrease in dielectric constant and grain size on La doping has been reported in BST [12] and BZT [13]. For the high dielectric constant capacitor applications, while the broadening of the phase transition has been achieved, the dielectric constant needs to be enhanced by increasing grain size and controlling oxygen vacancies. Indeed, the 3LBTS, whose microstructure comprises of a compact arrangement of relatively larger 2D (leafy) grains, shows a slight enhancement in the dielectric constant although the FWHM has also correspondingly decreased (Fig. 4 and Table 1).

We note that the temperature of the loss maxima as well as the dielectric loss peak value increases with increase in frequency. Similar behaviour was observed in La doped BZT [13] and La doped BTS [12]. The dielectric loss values at room temperature and 10 kHz are however higher than those in La doped BZT [13]. This could be due to conductive dielectric losses due to oxygen vacancies in Sn doped BT samples [12].

Polarization (P) verses Electric field (E) hysteresis loops at room temperature for all the samples are shown in Fig. 6. The remnant polarization and coercive field for BTS, 1LBTS, 2LBTS and 3LBTS are listed in Table 2. In pure BTS, some measure of saturation in polarization is observed, but full saturation was not obtained even after increasing E to 30 kV/cm. The PE loops for La doped samples do not show saturation indicating the presence of pinned polar nano-regions [12]. Since $T_m$ decreases with increase in La doping, it is expected that the PE loops obtained at fixed room temperature would also become slimmer with increase in La doping. However coercive field in 2LBTS is only marginally smaller than in 1LBTS. This might be attributed to enhanced grain boundary pinning due to small grain size in 2LBTS [29]. In addition, pinning due to oxygen vacancies arising due to oxygen loss might also play a role [14].

## 4. Conclusions

In conclusion, we report a detailed microstructural and dielectric characterization of a series of lanthanum doped barium stannate titanate ceramics. The samples were prepared by solid state reaction technique and XRD analysis indicated single phase microstructure. SEM images indicated that while a low La doping of 1 atomic % yielded inhibited grain growth in 1LBTS, higher amounts of La doping yield larger leafy grains. This peculiar grain growth and microstructure evolution can be attributed to the combination of the following two competing factors: (i) suppression of O vacancies due to La doping leading to grain growth inhibition and (ii) enhancement of mechanism leading to nucleation and growth of 2-D (leafy) grains for La doping higher than 2 atomic % due to eutectic phenomena. Dielectric permittivity measurements indicate that 1 atomic % of La enhances diffuse phase transition in $Ba(Ti_{0.80}Sn_{0.2})O_3$, but the frequency dependence of $T_m$ is observed only at 2 atomic % of La doping. This relaxor behaviour, the increase in the intensity of Raman peaks with increase in La doping, as well as the non-saturated PE hysteresis loops in La doped BTS indicate the presence of polar nano-regions. The decreased dielectric permittivity of La doped BTS can, at least partially, be attributed to the reduced grain size. The ferroelectric coercivity of 2LBTS is only marginally smaller in comparison to 1LBTS which could be explained on the basis of pinning due to grain boundaries and oxygen vacancies in 2LBTS. For the high dielectric constant capacitor applications, while the broadening of the phase transition has been achieved, the dielectric constant needs to be enhanced further by increasing the grain size and controlling oxygen vacancies.

## 5. Acknowledgements

Authors thank Dr. Satyendra Singh from SCNS, JNU, New Delhi, for access to PE loop tracer and Advance Instrument Research Facility AIRF, JNU, New Delhi, India, for access to SEM. B.B. acknowledges financial support via UGC start-up grant and DST Purse grant.

**Figure captions**

**FIGURE 1.** (Colour online) Room temperature XRD patterns of ceramic powder: (a) BTS (b) 1LBTS (c) 2LBTS (d) 3LBTS. Portion ($36 \leq 2\theta \leq 47$) of the XRD plot is shown magnified in inset.

**FIGURE 2.** Scanning electron microscopy (SEM) images of ceramics pellets: (a) BTS (b) 1LBTS (c) 2LBTS (d) 3LBTS.

**FIGURE 3.** (Colour online) Room temperature Raman spectra recorded from ceramics pellets: (a) BTS (b) 1LBTS (c) 2LBTS (d) 3LBTS.

**FIGURE 4.** (Colour online) Temperature dependent dielectric permittivity of ceramics pellets: (a) BTS (b) 1LBTS (c) 2LBTS (d) 3LBTS.

**FIGURE 5.** (Colour online) Plot of inverse dielectric constant at 100 kHz as function of temperature of ceramic pellets: (a) BTS (b) 1LBTS (c) 2LBTS (d) 3LBTS. Inset shows the corresponding $\ln(1/\epsilon - 1/\epsilon_m)$ verses $\ln(T-T_m)$. Linear fittings in the high temperature regime are indicated by solid lines.

**FIGURE 6.** (Colour online) Room temperature Polarization (P) verses Electric Field (E) hysteresis loop of ceramic pellets: (a) BTS, (b) 1LBTS, (c) 2LBTS, (d) 3LBTS.

**Table captions**

**Table 1.** The values of $T_m$, $\epsilon_m$ and $\gamma$ obtained from modified Curie–Weiss Law and FWHM values of the $\epsilon$ (T) at 100 kHz are given: (a) BTS, (b) 1LBTS, (c) 2LBTS, (d) 3LBTS.

**Table 2.** Coercive field ($E_c$) and remnant polarization ($P_r$) of ceramic pellets are given: (a) BTS, (b) 1LBTS, (c) 2LBTS, (d) 3LBTS.

Table 1

| La Doping | z=0.0 | z=0.01 | z=0.02 | z=0.03 |
|---|---|---|---|---|
| $T_m(K)$ | 238 | 204.8 | 146.8 | 121.5 |
| $\epsilon_m(F/m)$ | 7361 | 3727 | 1333 | 2182 |
| $\delta T_{relax} = T_m$ (1 MHz) —$T_m$ (1 kHz) | 0 | 4 | 10 | 16 |
| FWHM calculated ( at 100 kHz) | 36 | 85 | 115 | 90 |
| γ | 1.42 | 1.88 | 2.017 | 2.018 |

Table 2

| La doping | Remnant Polarization $P_r$ ($\mu C/cm^2$) | Coercive Electric Field $E_c$ (kV/cm) |
|---|---|---|
| z=0.00 | 0.70 | 3.92 |
| z=0.01 | 0.50 | 2.75 |
| z=0.02 | 0.147 | 2.71 |
| z=0.03 | 0.095 | 1.22 |

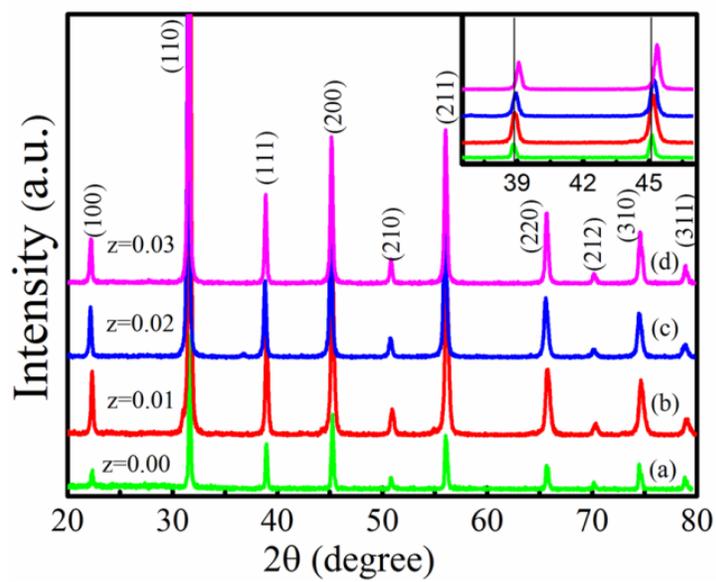

**FIGURE 1.**

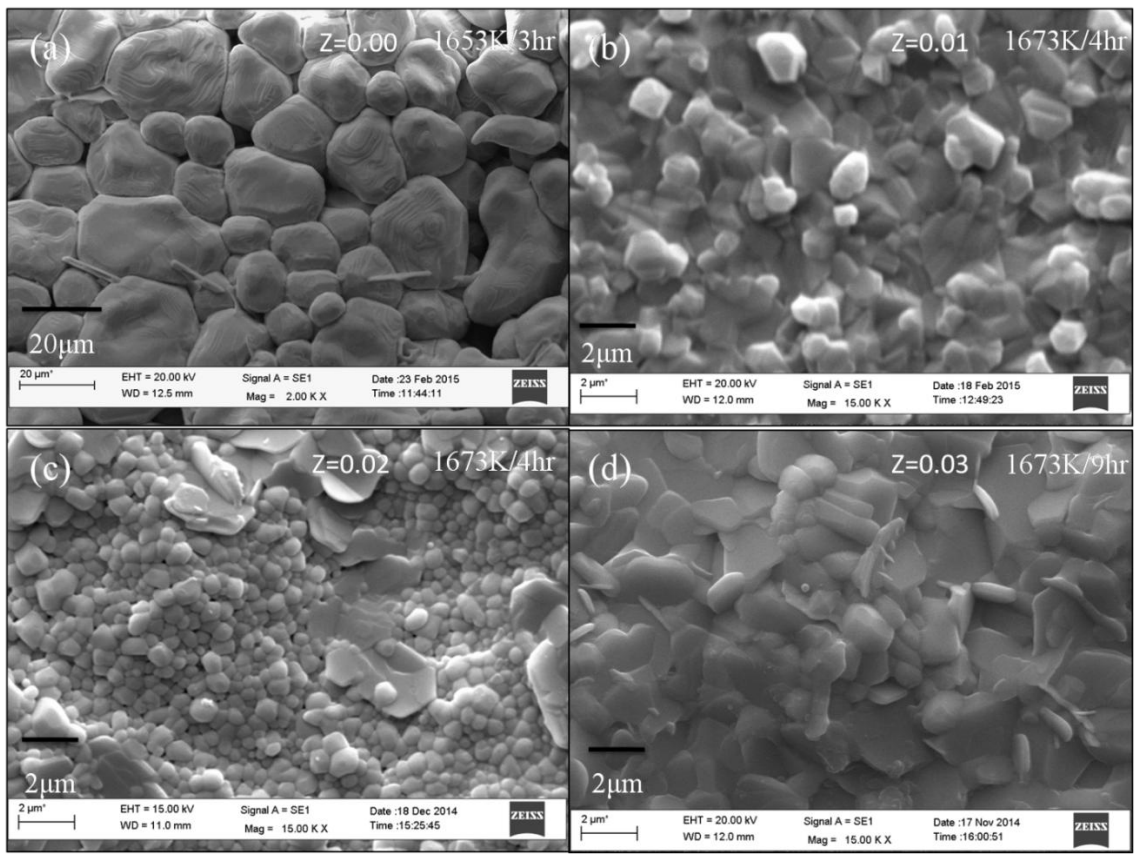

**FIGURE 2.**

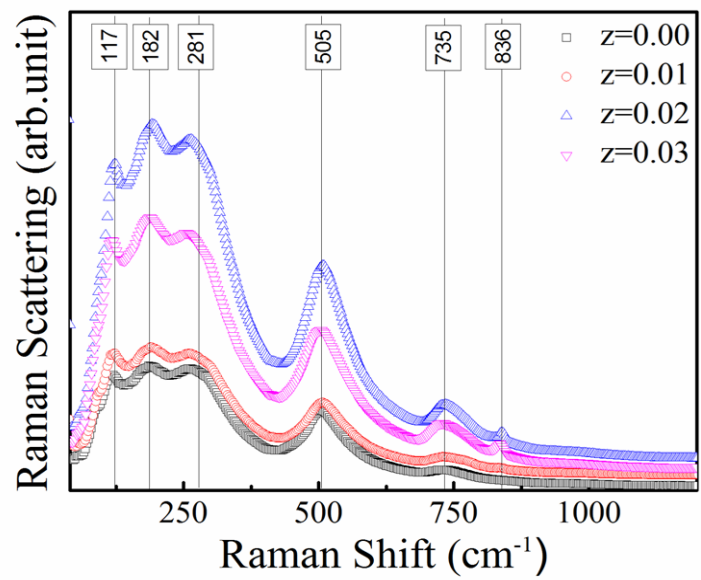

**FIGURE 3.**

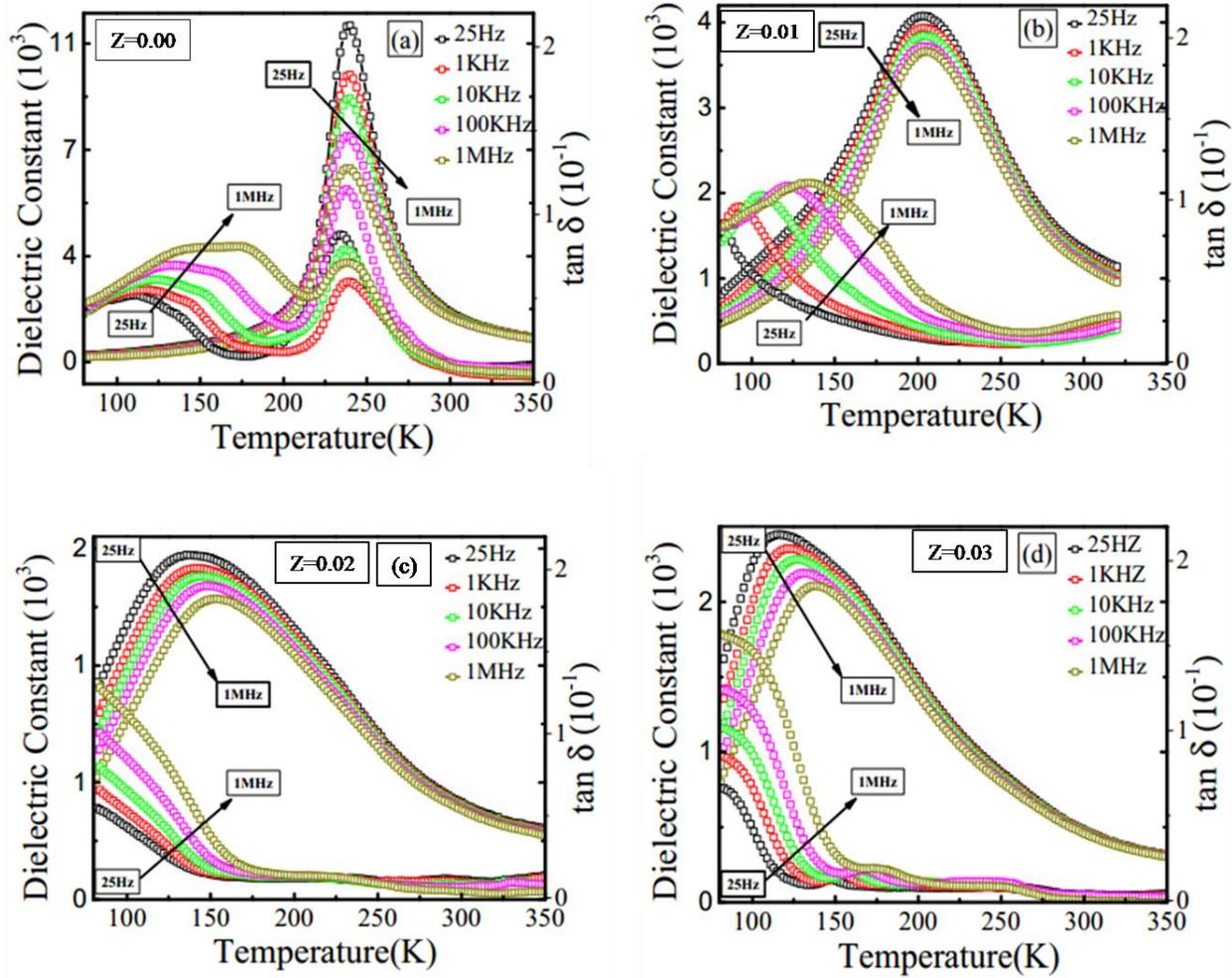

**FIGURE 4.**

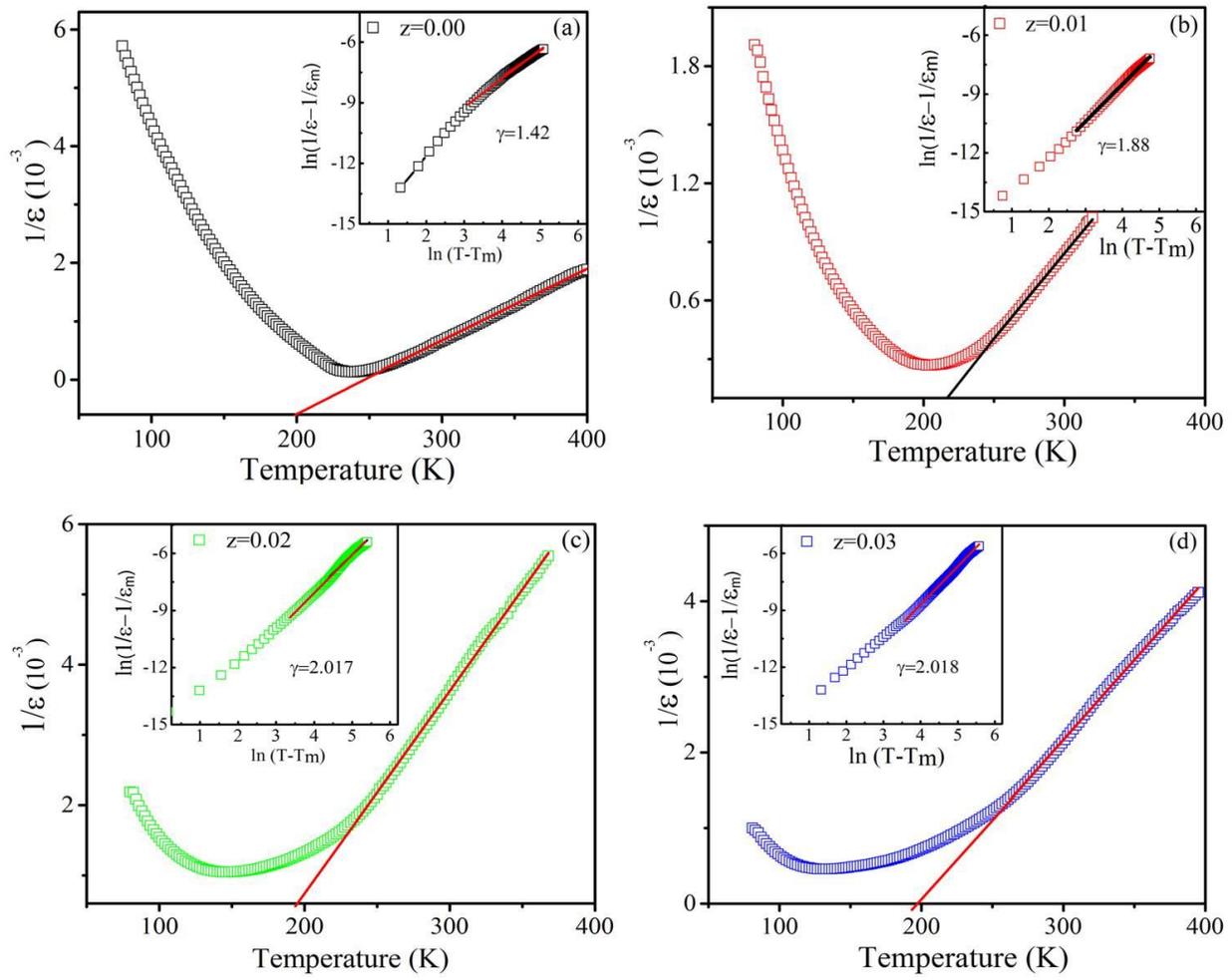

**FIGURE 5.**

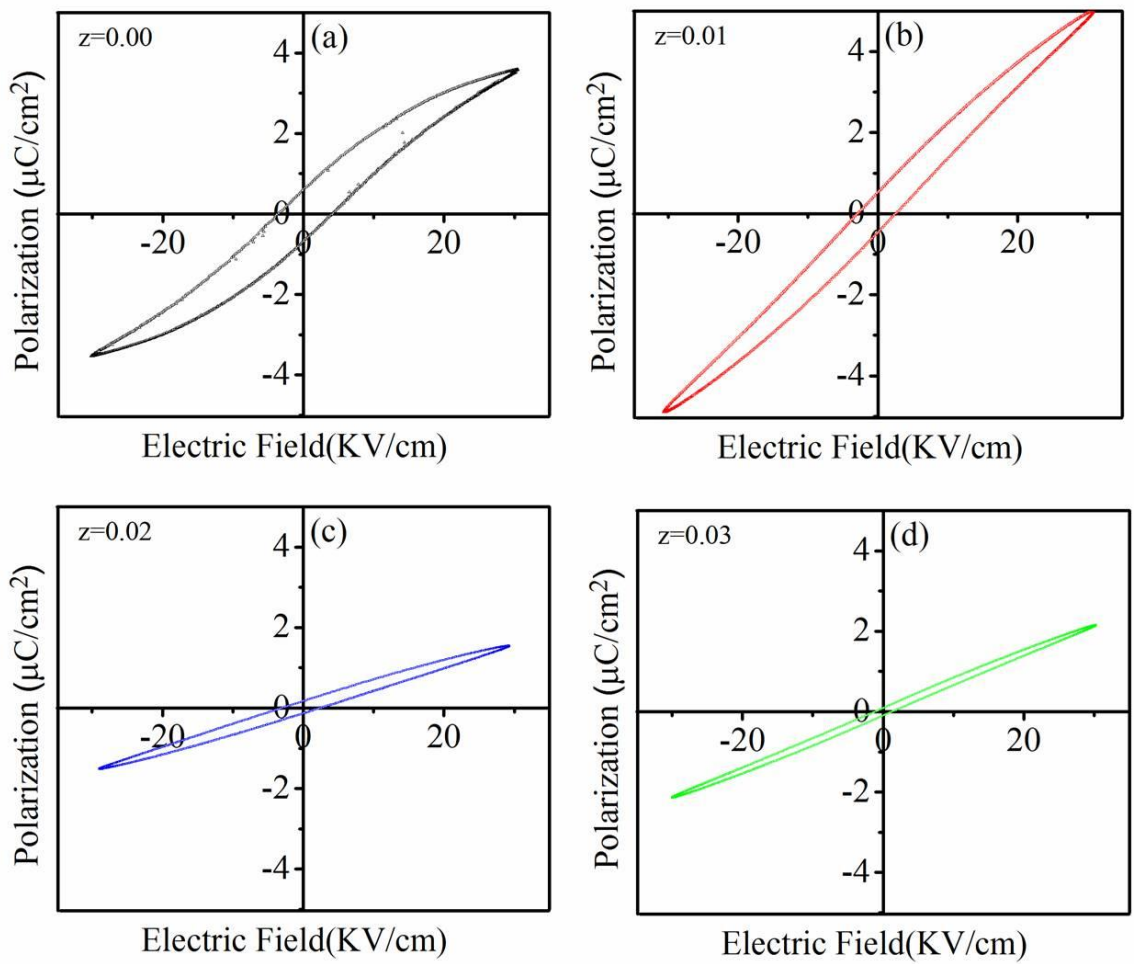

**FIGURE 6.**